\documentclass[preprint,onecolumn,aps,pre,eqsecnum,a4paper]{revtex4-1}
\usepackage[latin9]{inputenc}
\usepackage{color}
\usepackage[english]{babel}
\usepackage{amsmath}
\usepackage{graphicx}
\usepackage[unicode=true,pdfusetitle,
 bookmarks=false,
 breaklinks=true,pdfborder={0 0 0},backref=false,colorlinks=true]
 {hyperref}

\makeatletter
\numberwithin{equation}{section}
\numberwithin{figure}{section}
\numberwithin{table}{section}

\usepackage{graphicx}

\DeclareGraphicsExtensions{.png,.pdf,.eps}
\graphicspath{{.}}

\makeatother

\begin{document}


\title{Is the World Local or Nonlocal? Towards an Emergent Quantum Mechanics
in the 21\textsuperscript{st} Century}

\author{Jan \surname{Walleczek}\textsuperscript{1}}

\email[Corresponding author: ]{walleczek@phenoscience.com}

\author{Gerhard \surname{Grössing}\textsuperscript{2}}

\affiliation{\textsuperscript{1}Phenoscience Laboratories, Novalisstrasse 11,
10115 Berlin, Germany}

\affiliation{\textsuperscript{2}Austrian Institute for Nonlinear Studies, Akademiehof,
Friedrichstr.~10, 1010 Vienna, Austria\vspace{2cm}
}
\begin{abstract}
What defines an emergent quantum mechanics (EmQM)? Can new insight
be advanced into the nature of quantum \textit{nonlocality} by seeking
new links between \textit{quantum} and \textit{emergent} phenomena
as described by self-organization, complexity, or emergence theory?
Could the development of a \textit{future} EmQM lead to a unified,
\textit{relational} image of the cosmos? One key motivation for adopting
the concept of emergence in relation to quantum theory concerns the
persistent failure in standard physics to unify the two pillars in
the foundations of physics: quantum theory and general relativity
theory (GRT). The total contradiction in the foundational, metaphysical
assumptions that define orthodox quantum theory \textit{versus} GRT
might render \textit{inter}-theoretic unification impossible. On the
one hand, \textit{indeterminism} and \textit{non}-causality define
orthodox quantum mechanics, and, on the other hand, GRT is governed
by causality and \textit{determinism}. How could these two metaphysically-contradictory
theories ever be reconciled? The present work argues that metaphysical
contradiction necessarily implies physical contradiction. The contradictions
are essentially responsible also for \textit{the measurement problem}
in quantum mechanics. A common foundation may be needed for overcoming
the contradictions between the two foundational theories. The concept
of emergence, and the development of an EmQM, might help advance a
\textit{common foundation}~-- physical \textit{and} metaphysical~--
as required for successful \textit{inter-theory unification}.
\end{abstract}
\maketitle

\section{Introduction}

The question ``Is the world local or \textit{non}-local?'' has long
guided work in quantum foundations. At the latest, this started with
the introduction, by Einstein, Podolsky, and Rosen (EPR), of the first
precise \textit{metaphysical} \textit{definitions} in relation to
nonlocality as a concept (Einstein \textit{et al.}~\citep{Einstein.1935can}).
80 years on, that question~-- rather surprisingly~-- remains unanswered
still. On the one hand, there is no doubt any longer that EPR-type
nonlocal correlations can be observed in quantum experiments by observers
who are separated at \textit{space-like} distances. On the other hand,
the \textit{ontological} question remains wholly undecided of whether
these nonlocal \textit{observations} might imply the actual existence
of a ``\textit{nonlocal reality}''~-- \textit{not} merely in terms
of an \textit{operational metaphor} as in orthodox quantum theory.
The prospect of fundamentally ``real nonlocality'' was proposed,
for example, by de Broglie--Bohm theory (Bohm~\citep{Bohm.1952interpr1,Bohm.1952interpr2}).
Inspired by both Bohm's proposal (Bohm~\citep{Bohm.1952interpr1,Bohm.1952interpr2})
and the EPR argument (Einstein \textit{et al.}~\citep{Einstein.1935can}),
John Bell succeeded in proving that no quantum theory based on the
joint assumptions of \textit{reality} and \textit{locality} could
successfully reproduce the predictions that are yielded by \textit{orthodox},
i.e., \textit{operationalist} quantum mechanics (Bell~\citep{Bell.1964einstein}). 

The seminal proof of Bell's theorem left open, however, the extraordinary
possibility that reality might be~-- ontologically-speaking~-- \textit{nonlocal
in nature}. That possibility, which necessarily reaches beyond \textit{operationalist}
quantum theory, is pursued by what has become known as the \textit{ontological},
\textit{realist}, \textit{approach} to quantum mechanics (e.g., Bohm
and Hiley~\citep{Bohm.1993undivided}). The project of developing
an `emergent quantum mechanics' (EmQM) is usually placed in the context
of \textit{realist} approaches to quantum mechanics.

The implications of an EmQM are startling, however, when viewed through
the lens of the orthodox perspective: instead of finding~-- at reality's
deepest levels~-- absolute ``\textit{quantum randomness}'', a future
EmQM, including also de Broglie--Bohm theory, would find ``\textit{quantum
interconnectedness}'', e.g., possibly in the form of instantaneous
\textit{nonlocal influences} across the universe. For example, when
John Bell was asked what the meaning was of nonlocality, he answered
that \textit{nonlocality} ``\ldots{} \textit{means that what you
do here has immediate consequences in remote places}'' (Mann and
Crease~\citep{Mann.1988interviewBell}). What might the phenomenon
of `\textit{emergence}' offer towards a new understanding of nonlocality
in the deeper sense of Bell's ``\textit{immediate consequences}''~--
beyond the standard \textit{operationalism} of orthodox theory?

\section{Why `emergence' in quantum mechanics? }

One key motivation for adopting the concept of (irreducible) emergence
in relation to quantum theory concerns the much-debated failure to
unify the two pillars in the foundations of physics: quantum theory
and general relativity theory (GRT). Therefore, the long-term project
of inter-theory unification might be injected with fresh thinking
via the introduction into quantum mechanics of the concept of emergence.
Why might that be so?

\begin{figure}[!tbh]
\begin{centering}
\vspace{7mm}
\includegraphics[scale=0.35]{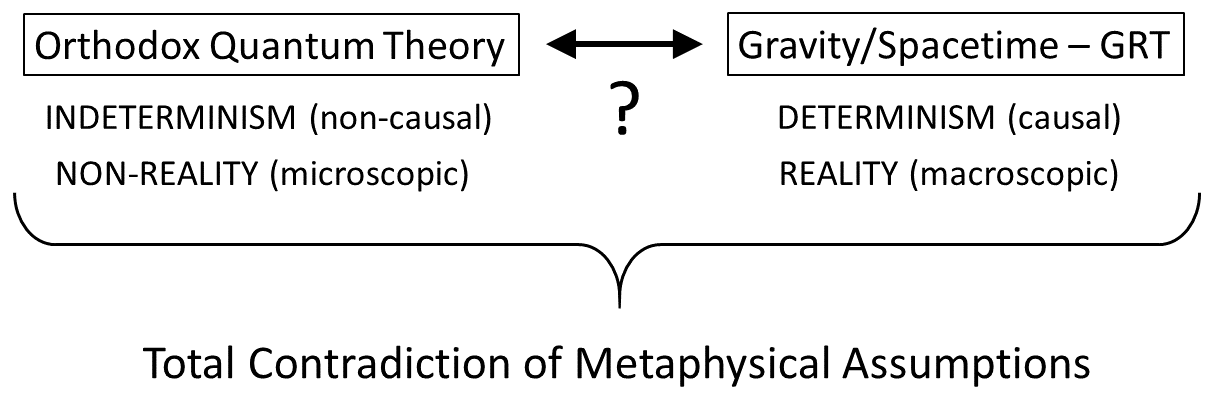}
\par\end{centering}

\caption{Total contradiction of metaphysical assumptions between orthodox quantum
theory and general relativity theory (GRT). \textit{Metaphysical}
contradiction implies \textit{physical} contradiction (see Sect.~2.1).
How is the reconciliation of metaphysical assumptions possible?{\small{}\label{fig:1}}}
\end{figure}

On the one hand, \textit{orthodox} quantum theory, as we understand
it today, is an entirely \textit{indeterministic} and \textit{non-causal}
theory, which presumes the \textit{complete absence} of any fundamental,
ontological reality at the level of the quantum. ``There is no quantum
world.'' Niels Bohr explained, ``There is only abstract quantum-mechanical
description'' (Petersen~\citep{Petersen.1963philNielsBohr}). On
the other hand, relativity theory (GRT) represents an ontological
theory of \textit{space-time reality}, in a decidely \textit{causal}
and \textit{deterministic} manner. Fig.~\ref{fig:1} illustrates
the fact that the metaphysical assumptions associated with the theories
contradict each other: ``indeterminism'' \textit{versus} ``determinism'',
and ``non-reality'' \textit{versus} ``reality''. These contradictions
are responsible also, of course, for the so-called \textit{measurement
problem} in quantum mechanics.

\subsection{Why `metaphysics' in quantum physics?}

Why is there this emphasis on \textit{metaphysical assumptions}? It
is helpful to remember that in the EPR argument already, which called
for the \textit{incompleteness} of orthodox quantum mechanics (Einstein
\textit{et al.}~\citep{Einstein.1935can}), it was the exact derivation
and definition of metaphysical assumptions which allowed the EPR argument
to have relevance to concrete problems facing quantum physics: Is
the world local or nonlocal? It was only through the consideration
of metaphysical notions like ``locality'', ``nonlocality'', ``causality'',
and ``reality'', that the breakthrough of Bell's theorem was possible
(Bell~\citep{Bell.1964einstein}). What is often lost in this picture
is the following: \textit{metaphysical} assumptions essentially constrain
the application of any \textit{mathematical theory} to concrete \textit{physical}
situations. Importantly, ``metaphysical'' is neither ``mystical''
nor ``irrational''. A metaphysical analysis refers to the first
principles and the foundational physical assumptions which inevitably
underpin \textit{any scientific} or \textit{mathematical analysis}
of nature. Often, foundational assumptions represent the \textit{preferred
world view} of the working scientist, including \textit{preconceived
notions} of what may, or may not, be possible in reality. Thus, by
adopting a new metaphysical position, a new vista might open up towards
the solution of a previously intractable scientific problem.

It appears likely that \textit{not} any amount of mathematical or
technical sophistication will reconcile the two theories~-- quantum
and relativity, \textit{unless the problem of their immediate metaphysical
opposition could be resolved also} (compare Fig.~\ref{fig:1}). Similarly,
any resolution of the measurement problem is likely to depend on the
``metaphysical reconciliation''~-- at the \textit{macroscopic}
and \textit{microscopic} levels~-- of any future \textit{physical
explanations}. Not suprisingly, it was John Bell~\citep{Bell.1987speakable-chapterSpeakable}
again who suggested ``\ldots{} that a real synthesis of quantum
and relativity theories requires not just technical developments but
radical conceptual renewal.''

\section{Towards an emergent quantum mechanics}

The research project of an EmQM follows the spirit of John Bell's
call for ``radical conceptual renewal'' (Bell~\citep{Bell.1987speakable-chapterSpeakable}),
a call consistent with his well-documented \textit{realist} expectations
about the future of quantum mechanics (e.g., Bell~\citep{Bell.2004nouvelle}).
EmQM research seeks a common foundation upon which might rest both
quantum theory \textit{and} GRT. Presently, the availability of a
common foundation is disputed or, at least, entirely unconfirmed.
However, the concept of `emergence' from self-organization, chaos,
or complexity, theory~-- once properly adapted~-- might offer a
\textit{universal framework}, both physically \textit{and} metaphysically,
for finally promoting ``\ldots{} a real synthesis of quantum and
relativity theories\ldots{}'' (Bell~\citep{Bell.1987speakable-chapterSpeakable}).

\begin{figure}[!tbh]
\begin{centering}
\includegraphics[scale=0.35]{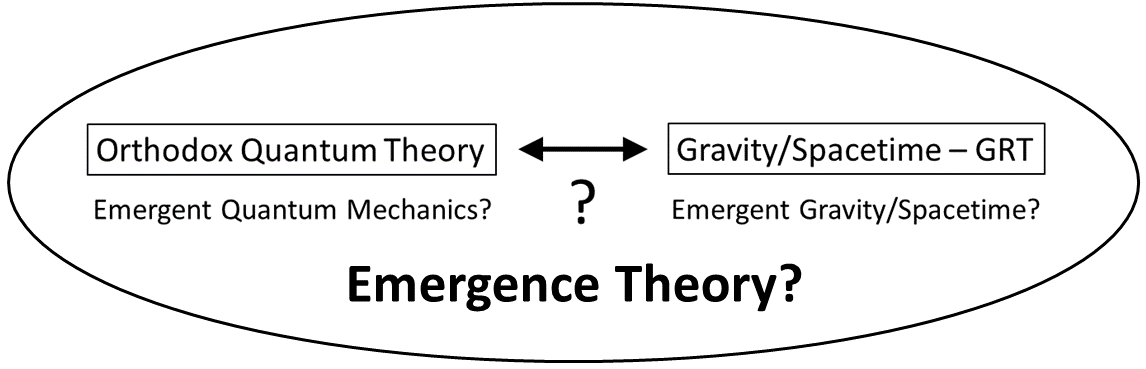}
\par\end{centering}

\caption{The concept of \textit{emergence} may provide a common \textit{physical}
and \textit{metaphysical foundation} in efforts to unify quantum and
relativity theories (GRT). A common foundation will be needed for
overcoming the deep metaphysical contradictions~-- in the \textit{orthodox}
approach~-- which have thus far prohibited success in inter-theory
unification (compare Fig.~\ref{fig:1}).{\small{}\label{fig:2}}}
\end{figure}

For some time now, the concept of emergence has found use already
in gravitational theory and in understanding the nature of space and
time. Both the puzzles and the possibilities of notions such as `emergent
gravity' and `emergent space-time' have been well summarized, for
example, by David Gross~\citep{Gross.2014quantum}: ``Many of us
are convinced that space is an emergent, not fundamental concept.
We have many examples of interesting quantum mechanical states, for
which we can think of some (or all) of the spatial dimensions as emergent.
Together with emergent space, we have the emergent dynamics of space
and thus emergent gravity. But it is hard to imagine how time could
be emergent? How would we formulate quantum mechanics without time
as a primary concept? Were time to be emergent, our understanding
of quantum mechanics would have to change.'' See Fig.~\ref{fig:2}
for a sketch illustrating the proposal that new understanding of quantum
mechanics, based on \textit{emergence}, could lessen, or even lift,
the inter-theoretic contradiction shown before in Fig.~\ref{fig:1}.

The key point is the following: once space-time and gravity are recast
in terms of fundamentally \textit{emergent} states or dynamics, this
invites the new view of the \textit{quantum} nature of reality in
terms of \textit{emergent dynamics} as well. Thus, a common conceptual
foundation might be developed~-- based on emergence as a guiding
principle~-- capable of bridging the vast chasm between quantum and
relativity theories. Maybe, then, there could be a new way to look
at the problem of inter-theory unification. In the future, there might
be theories describing some kind of ``emergent quantum gravity''
as a result. For example, pioneering work based upon a locally-deterministic
form of an ``emergent quantum mechanics'' was carried out by 't
Hooft~\citep{tHooft.2007emergent} (2007).

\section{What is emergence?}

\begin{figure}[!tbh]
\begin{centering}
\includegraphics[scale=0.3]{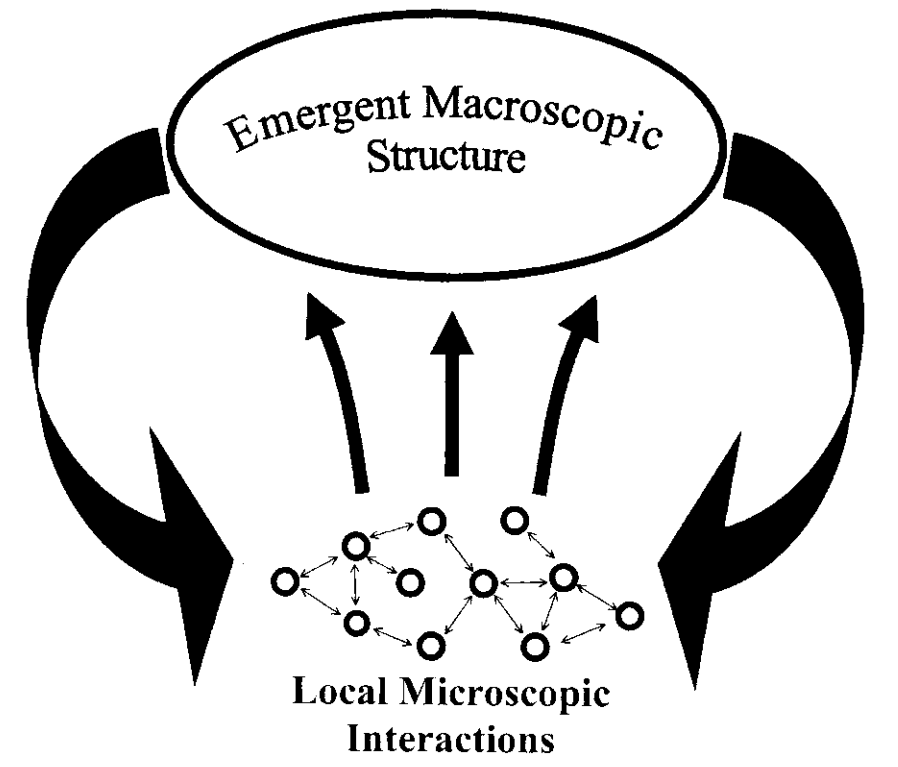}
\par\end{centering}

\caption{Illustration of self-referential, dynamical interactions across levels
of organization~-- \textit{microscopic} and \textit{macroscopic}
(from Walleczek~\citep{Walleczek.2000self-organized}). Both top-down
and bottom-up causal flows are indicated in the formation of an \textit{emergent
macroscopic structure}. Emergence describes the \textit{spontaneous
synchronization} of individual random motions into a unified collective
motion. Emergence accounts for the rise of \textit{global} macroscopic
order from \textit{local} microscopic randomness. An example is the
emergent formation of spatio-temporal, \textit{long-range coherence}.{\small{}\label{fig:3}}}
\end{figure}

In a more general context, what is \textit{emergence}? The concept
of emergence is present under the guise of many different names and
theories: complexity theory, chaos theory, self-organization theory,
non-linear dynamical systems theory, synergetics, cybernetics, fractal
sets, cellular automata, and so on. Emergent events are characterized
by sensitive dependencies on \textit{initial} conditions in combination
with \textit{evolving boundary} conditions. Generally, emergence accounts
for the rise of global macroscopic \textit{order} from local microscopic
\textit{randomness}. Both, top-down \textit{and} bottom-up causal
flows are implicated in the formation of an \textit{emergent macroscopic
structure} (Fig.~\ref{fig:3}). These causal flows are considered
to be \textit{relational} because vastly different levels in the hierarchy
of organization are \textit{actively interconnecting} without exclusive
priority of one level over another (see legend to Fig.~\ref{fig:3}).

\subsection{Determination without pre-determination: ``effective indeterminism''}

An important dimension in the development of an EmQM, i.e., for any
theory which connects (classical) emergence theory with quantum mechanics,
is the question of the inherently \textit{probabilistic} nature of
quantum phenomena. Crucially, in-principle \textit{unpredictability},
as well as \textit{uncontrollability}, of individual microscopic (quantum)
events must be ensured by any kind of \textit{non-orthodox} theory
which claims success in reproducing the predictions that are yielded
by \textit{orthodox} quantum mechanics. Otherwise, for example, the
\textit{non-signalling theorem} of quantum mechanics would be instantly
violated as we have discussed before at great length (Walleczek and
Grössing~\citep{Walleczek.2014non-signalling,Walleczek.2016nonlocal}).
Critical in this context is that emergent phenomena are subject to
unpredictability as a consequence of the \textit{intrinsically self-referential}
nature of the governing dynamics as illustrated in Fig.~\ref{fig:3}
(e.g., compare also the halting problem in computational theory).
A well-known example is the phenomenon of \textit{deterministic chaos},
which provides a vivid image of determination without predetermination,
i.e., ``effective indeterminism''. Future work in EmQM foundations
needs to clearly establish the limits and conditions under which such
scenarios apply in alternative models for quantum phenomena, including
for quantum nonlocality.

\section{Outlook: new approaches in realist quantum mechanics}

What are the prospects for an `Emergent Quantum Mechanics'? It is
possible~-- in principle~-- that the universe is deterministic,
e.g., \textit{nonlocally causal} in light of EPR-type nonlocal correlations.
Yet~-- at the same time~-- even a deterministic universe can have
an \textit{open future} in the context of emergence theory, i.e.,
a future where both the free-choice performances of an observer/agent,
and other physical processes in the cosmos, are \textit{not pre-determined}
by past events. As was explained in Sect.~4, emergent dynamical processes
are well-known for being governed by entirely deterministic relations,
and yet these very same processes can be \textit{without} pre-determined
outcomes in the future. As a consequence of the \textit{intrinsically
self-referential} nature of emergent phenomena, the \textit{in-principle
unpredictability} of \textit{individual} microscopic events is granted.
Whether such concepts might apply productively in a future quantum
mechanics remains for now a promising \textit{vision}. However, the
resurgence of interest in \textit{ontological} approaches to quantum
mechanics, including those pioneered and envisioned by David Bohm~\citep{Bohm.1952interpr1,Bohm.1952interpr2}
and John Bell~\citep{Bell.1976theory,Bell.1977free,Bell.1987speakable-chapterSpeakable,Bell.2004nouvelle}
may further increase interest in the project of an EmQM (e.g., see
also Bohm and Hiley~\citep{Bohm.1993undivided}; Holland~\citep{Holland.1993}).

In conclusion, a new wave of work has drawn attention to ontological,
realist questions in quantum mechanics: Does the concept of `nonlocality'
reflect the true nature of reality? Is the quantum state real? Is
the wave function $\psi$ a reality? On the \textit{theoretical} side,
especially work by Harrigan and Spekkens~\citep{Harrigan.2010einstein}
has renewed interest in ontological theory, including de Broglie--Bohm
theory, by presenting the productive distinction between $\psi$-ontic
and $\psi$-epistemic approaches to quantum mechanics. In that context,
our own recent work showed that \textit{nonlocal quantum information
transfers}, which are \textit{inevitably} associated with any $\psi$-ontic
quantum theory, including Bohm's theory, need \textit{not} violate
the \textit{non-signalling theorem} (Walleczek and Grössing~\citep{Walleczek.2016nonlocal}).
On the \textit{experimental} side, the important work by Kocsis \textit{et
al.}~\citep{Kocsis.2011observing}, Ringbauer \textit{et al.}~\citep{Ringbauer.2015measurements},
and Mahler \textit{et al.~}\citep{Mahler.2016experimental}, has
advanced fresh insight into the \textit{non-orthodox} option of nonlocality
as a \textit{reality}, e.g., the reality of the wave function $\psi$.
Finally, the most recent available findings provide a ``compelling
visualization''~-- as the authors put it~-- ``of the nonlocality
inherent in any realistic interpretation of quantum mechanics'' (Mahler
\textit{et al.~}\citep{Mahler.2016experimental}).
\begin{acknowledgments}
Work by Jan Walleczek at Phenoscience Laboratories (Berlin) is partially
funded by the Fetzer Franklin Fund of the John E. Fetzer Memorial
Trust. Work by Gerhard Grössing at the Austrian Institute for Nonlinear
Studies (Vienna) is also partially funded by the Fetzer Franklin Fund
of the John E. Fetzer Memorial Trust. The authors wish to thank Siegfried
Fussy, Johannes Mesa Pascasio, Herbert Schwabl and Nikolaus von Stillfried
for their excellent contributions.
\end{acknowledgments}

\providecommand{\href}[2]{#2}\begingroup\raggedright\endgroup

\end{document}